\documentclass[useAMS,usenatbib]{mn2e}
\usepackage{graphicx}
\usepackage{aas_macros}

\title[Space density of IPs from \emph{Swift}-BAT]{Constraints on the space density of intermediate polars from the \emph{Swift}-BAT survey}

\author[M.L. Pretorius and K. Mukai]{Magaretha L. Pretorius,$^{1}$\thanks{E-mail: retha.pretorius@astro.ox.ac.uk (MLP); koji.mukai@nasa.gov (KM)} and Koji Mukai$^{2,3}$\footnotemark[1]\\
$^{1}$Department of Physics, University of Oxford, Denys Wilkinson Building, Keble Road, Oxford OX1 3RH, United Kingdom\\
$^{2}$CRESST and X-ray Astrophysics Laboratory, NASA/GSFC, Greenbelt, MD 20771, USA\\
$^{3}$Department of Physics, University of Maryland, Baltimore County, 1000 Hilltop Circle, Baltimore, MD 21250, USA
\\}

\begin{document}


\pagerange{\pageref{firstpage}--\pageref{lastpage}} \pubyear{}

\maketitle

\label{firstpage}

\begin{abstract}
We construct a complete, hard X-ray flux-limited sample of intermediate polars (IPs) from the \emph{Swift}/BAT 70-month survey, by imposing selection cuts in flux and Galactic latitude ($F_X > 2.5 \times 10^{-11}\,\mathrm{erg\,cm^{-2}s^{-1}}$ at 14--195~keV, and $|b|>5^\circ$). We then use it to estimate the space density ($\rho$) of IPs. Assuming that this sample of 15 long-period systems is representative of the intrinsic IP population, the space density of long-period IPs is $1^{+1}_{-0.5} \times 10^{-7}\,\mathrm{pc^{-3}}$. The \emph{Swift}/BAT data also allow us to place upper limits on the size of a hypothetical population of faint IPs that is not included in the flux-limited sample. While most IPs detected by \emph{Swift}/BAT have 14--195~keV luminosities of $\sim 10^{33} {\rm erg s^{-1}}$, there is evidence of a fainter population at $L_X \sim 10^{31} {\rm erg s^{-1}}$. We find that a population of IPs with this luminosity may have a space density as large as $5\times 10^{-6}\,\mathrm{pc^{-3}}$. Furthermore, these low-luminosity IPs, despite appearing rare in observed samples, are probably at least as intrinsically common as the brighter systems that are better represented in the known IP sample. 
\end{abstract}

\begin{keywords}
binaries -- stars: cataclysmic variables, -- X-rays: binaries -- methods: observational, statistical.
\end{keywords}

\section{Introduction}
\label{sec:intro}
Intermediate polars (IPs) are a class of cataclysmic variables (CVs) in which a magnetic white dwarf accretes material from a late-type, Roche-lobe filling binary companion. The magnetic field of the white dwarf in an IP is sufficiently strong to control the inner accretion flow, but not to synchronize the rotation of the white dwarf with the binary orbit.

Measurements of properties of the Galactic CV population, such as the space densities ($\rho$) of different classes of CVs, are important in constraining models of CV formation and evolution. 
IPs are more generally of interest, because, despite being the smallest observed class of CVs (there are 118 IPs and IP candidates, out of a total of 1166 CVs, in version 7.20 of \citealt{rkcat}), it is possible that they dominate Galactic X-ray source populations above $L_X \sim 10^{31}\,{\rm erg s^{-1}}$ in the 0.5--8.0~keV band (see e.g.\ \citealt{Muno09,mcvrho}). 

We have recently provided space density measurements for IPs and polars (magnetic CVs with white dwarfs locked in synchronous rotation with the binary orbit), based on the \emph{ROSAT} Bright Survey \citep{mcvrho}. Polars are luminous soft X-ray sources, and even this bright survey yielded a relatively large sample (24 systems). However, the \emph{ROSAT} sample contained only 6 IPs (and 2 of those were unusual short-period systems), leading to a relatively less robust $\rho$ measurement for IPs.

The Burst Alert Telescope (BAT) on board \emph{Swift} has been observing for about 10 years in the 14--195~keV band (compared to the much softer 0.5--2.0~keV \emph{ROSAT} band).
IPs have hard X-ray spectra, often modelled as intrinsically absorbed thermal bremsstrahlung (e.g.\ \citealt{Patterson94}). The usually high intrinsic absorption, as well as the hard unabsorbed SED itself, means that most known IPs are faint soft X-ray sources.
IPs are, however, making up a significant fraction of the Galactic sources 
detected at higher energies, e.g., in the \emph{INTEGRAL}/IBIS and \emph{Swift}/BAT surveys (see e.g.\ \citealt{bat70month}; \citealt{Bird10}), and these surveys have significantly increased the known IP sample (e.g.\ \citealt{Scaringi10}). Although the IBIS and BAT instruments observe in similar energy bands, the \emph{Swift}/BAT survey exposure is more uniform over the sky \citep{bat70month}, implying that it is better suited to defining flux-limited samples of objects than the \emph{INTEGRAL}/IBIS survey.

Here we use the \emph{Swift}-BAT 70-month survey to construct a well-defined IP sample, and use it to constrain the space density of IPs. In Section~\ref{sec:sample}, we present the hard X-ray flux-limited IP sample, together with distance ($d$) and X-ray luminosity ($L_X$) estimates. In Section~\ref{sec:rhocalc}, we describe the space density calculation, and present our results. Finally, we discuss the results and summarize the conclusions in Sections~\ref{sec:discussion} and \ref{sec:conclu}.

\section{The hard X-ray flux-limited IP sample}
\label{sec:sample}
Since it is not possible at the moment to construct a usefully large volume-limited CV sample, a purely flux-limited sample is the most suitable for our purpose (e.g.\ \citealt{PretoriusKniggeKolb07,halpha1,halpha2}). A total of 1210 sources were detected in the \emph{Swift}-BAT 70-month survey; most of the Galactic sources are interacting binaries (with 59 being known CVs\footnote{\cite{bat70month} classify 55 sources as CVs. We include also V1062 Tau, AH Men, V2487 Oph, and V603 Aql, which were counted as novae or nova-likes, rather than CVs, in that work.}), and 65 sources were of unknown type \citep{bat70month}.
Most of the CVs (34 systems) are IPs, or good IP candidates.
The survey has reached a flux limit of $1.03 \times 10^{-11}\,\mathrm{erg\,cm^{-2}s^{-1}}$ over 50\% of the sky and $1.34 \times 10^{-11}\,\mathrm{erg\,cm^{-2}s^{-1}}$ over 90\% of the sky \citep{bat70month}.

\subsection{Constructing a complete IP sample}
\begin{figure}
 \includegraphics[width=84mm]{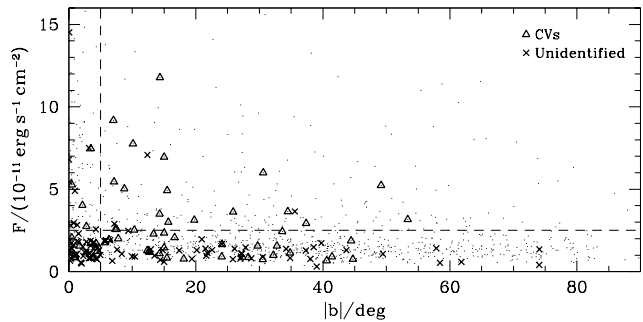} 
  \caption {Hard X-ray flux plotted against Galactic latitude for sources detected in the \emph{Swift}-BAT 70-month survey. The small points are all sources in the catalogue. CVs are over-plotted as triangles, and sources classified as ``Unknown'' are shown as crosses. The dashed lines indicate the selections cuts of $F_X > 2.5 \times 10^{-11}\,\mathrm{erg\,cm^{-2}s^{-1}}$ and $|b|>5^\circ$, which leave 221 objects securely identified as AGN and other object types that are not of interest in this work, 19 CVs, and only 3 unclassified sources (these can be excluded as CVs; see text). The CV sample contains 15 IPs, all with secure classification.}
 \label{fig:cuts}
\end{figure}

Our goal is to define a sample that includes as many IPs as possible, but no unidentified sources, by applying simple, well-defined selection cuts to the \emph{Swift}-BAT 70-month catalogue. 
In Fig.~\ref{fig:cuts}, we plot the BAT flux against Galactic latitude of all sources in the 70-month catalogue. The dashed lines in the plot show the selection cuts we choose, namely $F_X > 2.5 \times 10^{-11}\,\mathrm{erg\,cm^{-2}s^{-1}}$ and $|b|>5^\circ$. This flux limit is sufficiently bright that the whole sky has been surveyed to that depth (see figure 10 of \citealt{bat70month}). A total of 243 sources satisfy these 2 selection cuts. Of those, 19 are CVs, and 3, namely SWIFT J1706.6-6146, SWIFT J1508.6-4953, and SWIFT J0451.5-6949, were classified ``Unknown''. 

All 3 unclassified sources can confidently be excluded as CVs. SWIFT J1706.6-6146, also called IGR J17062-6143, is a hard X-ray transient, identified as a type-I X-ray burst from an X-ray binary \citep{Degenaar13}. SWIFT J1508.6-4953 is identified with the radio source PMN J1508-4953, which is far too bright to be a CV ($\ga 0.6\,{\rm Jy}$ at $8\,{\rm GHz}$; \citealt{Massardi08}), and is likely a flat spectrum radio quasar \citep{Tuerler12}. Finally, SWIFT J0451.5-6949 has been recognized as a high-mass X-ray binary in the LMC (e.g.\ \citealt{Bartlett13}).

Of the 19 CVs, 15 are IPs, all above the orbital period ($P_{orb}$) gap (the clear dip in the period distribution of CVs, between roughly 2 and 3~h). The other 4 CVs (BY Cam, V1432 Aql, AM Her, and SS Cyg) are well-studied systems, and clearly not IPs. Our selection cuts thus lead to a complete flux-limited sample of 15 IPs, which we list in Table~\ref{tab:distances}.

\begin{table*}
 \centering
 \begin{minipage}{168mm}
  \caption{The 15 IPs detected above $F_X = 2.5 \times 10^{-11}\,\mathrm{erg\,cm^{-2}s^{-1}}$ and $|b|=5^\circ$ in the \emph{Swift}-BAT 70-month survey, together with their orbital and white dwarf spin periods, 14--195~keV flux and 90\% confidence interval, distance estimates, and 14--195~keV hard X-ray luminosities. The best estimate of the fraction of the total IP space density contributed by each system ($\rho_j/\rho_0$; see Section~\ref{sec:rhocalc}) is given in the 7th column. References are for orbital and spin periods, and published distances, where available.}
  \label{tab:distances}
  \begin{tabular}{@{}lllllllll@{}}
  \hline
System  & $P_{orb}/\mathrm{h}$ & $P_{spin}/\mathrm{s}$ & $F_X/(10^{-11}\,\mathrm{erg\,cm^{-2}s^{-1}})$  & $d/\mathrm{pc}$ & $\mathrm{log}(L_X/\mathrm{erg\,s^{-1}})$ & $\rho_j/\rho_0$ & References \\
 \hline
V1223 Sgr        &  3.366 &  746   & 11.8 (11.4--12.1) &  527 (+54, -43)    & 33.6 (0.1)        & 0.019 & 1,2,3   \\  
V2400 Oph        &  3.41: &  927.6 &  5.0 (4.7--5.4)   &  280 (+150, -100)  & 32.7 (0.4)        & 0.130 & 4       \\  
AO Psc           &  3.591 &  805.2 &  3.2 (2.9--3.5)   &  330 (+180, -120)  & 32.6 (0.4)        & 0.150 & 5,6,7   \\  
IGR J16500-3307  &  3.617 &  571.9 &  2.6 (2.2--3.0)   &  430 (+240, -150)  & 32.8 (0.4)        & 0.107 & 8,9     \\  
V405 Aur         &  4.264 &  545.5 &  3.5 (3.2--3.8)   &  380 (+210, -130)  & 32.8 (0.4)        & 0.100 & 10,11,12\\  
FO Aqr           &  4.862 & 1254.3 &  5.2 (4.9--5.6)   &  450 (+240, -160)  & 33.1 (0.4)        & 0.049 & 13,14   \\  
PQ Gem           &  5.193 &  833.4 &  3.1 (2.8--3.5)   &  510 (+280, -180)  & 33.0 (0.4)        & 0.062 & 15,16,17\\  
TV Col           &  5.486 & 1909.7 &  6.0 (5.7--6.3)   &  368 (+17, -15)    & 33.0 (0.4)        & 0.062 & 18,19,20\\  
IGR J15094-6649  &  5.890 &  809.4 &  2.6 (2.3--2.9)   &  960 (+530, -340)  & 33.5 (0.4)        & 0.024 & 8,21    \\  
XY Ari           &  6.065 &  206.3 &  3.6 (3.2--4.1)   &  270 (100)         & 32.5 (+0.3, -0.4) & 0.199 & 22,23,24\\  
EI UMa           &  6.434 &  745.7 &  2.9 (2.6--3.2)   &  750 (+100, -200)  & 33.3 (+0.1, -0.3) & 0.033 & 25,26   \\  
NY Lup           &  9.86  &  693.0 &  9.2 (8.7--9.6)   &  680 (+170, -130)  & 33.7 (0.2)        & 0.015 & 27      \\  
V1062 Tau        &  9.982 & 3685   &  2.5 (2.1--3.0)   & 1400 (+700, -500)  & 33.8 (+0.3, -0.4) & 0.014 & 28,29   \\  
V2731 Oph        & 15.42  &  128.0 &  6.9 (6.5--7.4)   & $>1000$            & $>33.9$          &$<0.011$& 30,31   \\  
GK Per           & 47.923 &  351.3 &  7.8 (7.4--8.1)   &  477 (+28, -25)    & 33.3 (0.05)       & 0.031 & 32,33,34\\  
 \hline
 \end{tabular}
\\
References: 
1. \cite{BeuermannHarrisonMcArthur04}; 
2. \cite{Osborne1985};
3. \cite{JablonskiSteiner1987};
4. \cite{Buckley1995};
5. \cite{Patterson2001};
6. \cite{Johnson2006};
7. \cite{Patterson84};
8. \cite{Pretorius09};
9. \cite{Bernardini12};
10. \cite{Haberl1994};
11. \cite{Still98};
12. \cite{Piirola08};
13. \cite{deMartino1994};
14. \cite{Patterson98};
15. \cite{Patterson94};
16. \cite{Hellier97};
17. \cite{Evans06};
18. \cite{Dai10};
19. \cite{Rana04};
20. \cite{McArthur01};
21. \cite{Butters09};
22. \cite{Littlefair2001};
23. \cite{Allan96};
24. \cite{Kamata91};
25. \cite{Thorstensen86};
26. \cite{Reimer08};
27. \cite{deMartino2006};
28. \cite{Thorstensen2010};
29. \cite{Hellier02};
30. \cite{deMartino2008}; 
31. \cite{Gansicke2005};
32. \cite{Harrison2013};
33. \cite{Crampton86};
34. \cite{Hellier04}.
\end{minipage}
\end{table*}

\subsection{Distance and X-ray luminosity estimates}
\label{sec:dist}
In order to measure a space density, we need distance estimates. Published distances are available for several systems, and for the remainder, we estimate distances using mid-IR photometry.

Trigonometric parallaxes have been measured for V1223 Sgr, TV Col, and GK Per \citep{BeuermannHarrisonMcArthur04,McArthur01,Harrison2013}. For 4 more systems, XY Ari, EI UMa, NY Lup, and V1062 Tau, distance estimates from photometric parallax of the donor star have been reported \citep{Littlefair2001,Reimer08,deMartino2006,Thorstensen2010}. 

For the rest of the systems in the sample, with the exception of V2731 Oph, we use the method outlined in \cite{mcvrho} to find distance estimates. In doing this, we assume that none of these systems has an evolved donor star. We use in all cases {\it WISE} photometry and the semi-empirical donor sequence of \cite{Knigge06}, and assume that the donor contributes $\simeq 30\%$ of the flux in the $W1$ band (estimated in Pretorius et al., in preparation). We then find the probability distribution function for the distance to each source, assuming Gaussian errors in predicted absolute magnitudes and the observed apparent magnitudes, and neglecting interstellar extinction, since it has a small effect in the mid-IR. 
There are no distance measurements in the literature for V2400 Oph, IGR J16500-3307, or IGR J15094-6649 to compare to, but for AO Psc, V405 Aur, FO Aqr, and PQ Gem, we find distances that are consistent with published estimates based on other techniques (see \citealt{Patterson84,Haberl1994,deMartino1994,Patterson94}).

V2731 Oph (also known as 1RXS J173021.5-055933) is a very long-period system ($P_{orb}\simeq 15\,{\rm h}$). \cite{Knigge06} does not predict absolute magnitudes for CV donors at such long periods (indeed, the population is likely dominated by CVs with evolved donors at periods beyond about 6~h; e.g.\ \citealt{Knigge06,KBP11}). \cite{deMartino2008} obtain a lower distance limit of between 1 and 1.6~kpc, using the donor spectral type range G0 to G6V \citep{Gansicke2005}. This makes V2731 Oph the most luminous system in our sample, and hence the one that makes the smallest contribution to $\rho$ ($\rho_j/rho \la 0.01$; see Section~\ref{sec:rhocalc}). 
For the space density calculation, we will simply assume a distance distribution that is Gaussian in $\log (d)$, with a median at $1900$~pc, and $\sigma_{log(d)}$ chosen so that the 16th percentile corresponds to $1000$~pc. Although this assumed distance distribution is not correct, it is justified in this context, since V2731 Oph does not significantly affect our space density estimate. The difference in the space density we would obtain with 2 extreme assumptions, namely excluding this system, or assuming the minimum allowed distance of 1 kpc, is $\simeq$1\%, which is negligible compared to the uncertainty in the estimate (see Section~\ref{sec:results}).

The distance and resulting $L_X$ estimates (for which we neglect interstellar photoelectric absorption, since it does not affect hard X-rays) are listed in Table~\ref{tab:distances}. Several of the distances are uncertain to $\simeq 50\%$, leading to very poorly constrained luminosities. 
The X-ray fluxes and luminosities are in the 14--195~keV band. Fluxes are based on a power-law spectral fit, and errors on flux are expressed as a 90\% confidence interval \citep{bat70month}, while the distance and ${\rm log}(L_X)$ estimates we give are the median, together with the 1-$\sigma$ confidence interval corresponding to the 16th and 84th percentile points.

\section{Calculating the space density}
\label{sec:rhocalc}
Here, we briefly describe the calculations we do to obtain a space density estimate with its uncertainty, as well as constraints on the space density of possible undetected populations. We then present the results.

\subsection{The method}
\label{sec:calc}
In both simulations described below, we assume that space density drops off exponentially with height above the Galactic plane ($z$), i.e. $\rho(z)=\rho_0 \mathrm{e}^{-|z|/h}$, and neglect the weaker radial dependence. 
We assume a single scale height of $h=120\,\mathrm{pc}$ for all IPs. This scale height is appropriate for a population with an age of $\simeq 10^{8.4}$~yr \citep{RobinCreze86}; our sample contains only long-period systems, and $\sim10^{8.4}$~yr is a reasonable approximation of a representative, average age for these CVs \citep{HowellNelsonRappaport01}. It is sufficient for our purposes to assume that long-period CVs are a single-age population (since systems evolve down to the top of the period gap in a short fraction of their total lives; this also means that the $|b|$ cut does not introduce a period-dependent bias)\footnote{We checked whether the $z$-distribution of the observed systems is consistent with this model. In a flux- and $|b|$-limited sample, the observed $z$-distribution is not expected to be the same as the intrinsic distribution; the effect of the selection cuts depends on the intrinsic $L_X$-distribution. We assume an intrinsic $L_X$-distribution that is Gaussian in $\log(L_X)$, distribute a population of systems in the model Galaxy, and then apply our flux- and $|b|$-cuts to simulate an observed sample. 
Using a Kolmogorov-Smirnov (KS) test, we find that the $z$-distribution of our sample is consistent with a model scale-height in the range $100$ to $300\,\mathrm{pc}$, for choices of intrinsic $L_X$-distributions that also produce simulated $L_X$-distributions consistent with that of our sample.}. 
We neglect interstellar absorption throughout, since our sample is based on hard X-ray data.

\subsubsection{The observed population}
\label{sec:calcobs}
Calculating $\rho$ simply involves counting the systems detected inside the volume observed by the survey. In order to find the survey volume, we use the relation 
\begin{equation}
V_j=\Omega \frac{h^3}{|\sin b|^3}\left[2-\left(x_j^2+2x_j+2\right)\mathrm{e}^{-x_j}\right]
\label{eq:V}
\end{equation}
(e.g.\ \citealt{TinneyReidMould93}), which accounts for both the flux- rather than volume-limited nature of the sample, and the dependence of $\rho$ on position in the Galaxy. 
The index $j$ represents the CVs in our sample, $\Omega$ is the solid angle covered by the survey, and $x_j=d_{max,j} |\sin b|/h$, with $d_{max,j}$ the maximum distance at which the system could have been detected, given its luminosity and the survey flux limit (i.e., $d_{max,j}=d_j\sqrt{F_{X,j}/F_{lim}}$).
The mid-plane space density is then the sum of the space densities represented by each system, i.e., $\rho_0=\sum_j 1/V_j$. The best-estimate fractional contribution that each system in our sample makes to the total space density ($\rho_j/\rho_0$) is given in the 7th column of Table~\ref{tab:distances}.

In order to find the error on $\rho_0$, we compute its probability distribution function using a Monte Carlo simulation that finds $\rho_0$ (as described above) for a large number of mock samples with properties that fairly sample the parameter space allowed by the data. The mock samples are generated by drawing a distance and flux for each of the 15 observed system from the appropriate distributions. These values are used to calculate each $1/V_j$, which is then weighted by a factor $\mu_j$, drawn from the probability distribution of the number of sources belonging to the population that corresponds to a particular observed system, that one expects to detect in the survey (see \citealt{NEPrho,nonmagphi,mcvrho}). We find that the error in $\rho_0$ is dominated by uncertainty in distances.

\subsubsection{A possible unobserved population}
\label{sec:calclim}

The most important assumption of the simulation described above is that the whole IP luminosity function is represented in the observed IP sample, including the faintest IPs in the intrinsic population. In other words, the space density calculation is only valid if IPs have $L_X \ga 10^{32} {\rm erg s^{-1}}$ in the BAT energy band (see Table~\ref{tab:distances}). A fainter population can be completely unrepresented in a flux-limited survey, since the survey volume is smaller for fainter $L_X$ (see \citealt{NEPrho,nonmagphi,mcvrho}). However, the non-detection nonetheless allows us to constrain the space density of a hypothetical faint populations of IPs.

We perform another Monte Carlo simulation, in which we distribute a model single-$L_X$ population of IPs in the model Galaxy, and then find the value of $\rho_0$ for which the predicted number of systems detected in the \emph{Swift}-BAT survey, after applying our $F_X$- and $|b|$-cuts, is 3 (a non-detection is then a 2-$\sigma$ result). We sample a range in assumed $L_X$ for the undetected population, below the faintest luminosities in the observed sample (from $10^{30}$ to $10^{32} {\rm erg s^{-1}}$).

\subsection{Results}
\label{sec:results}

\subsubsection{The probability distribution function of $\rho_0$}
\label{sec:rhoresult}

\begin{figure}
 \includegraphics[width=84mm]{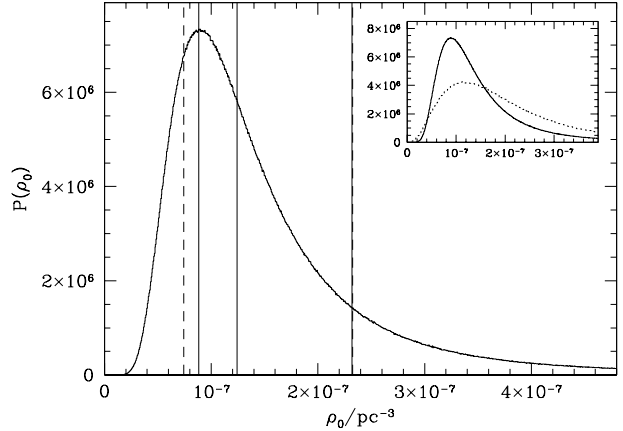} 
  \caption{The probability distribution function of $\rho_0$, resulting from the simulation described in Section~\ref{sec:calcobs}. The solid vertical lines are at the mode, median, and mean ($8.9 \times 10^{-8}$, $1.2 \times 10^{-7}$, and $2.3 \times 10^{-7}\,\mathrm{pc^{-3}}$). The dashed lines mark a 1-$\sigma$ interval from $7.4 \times 10^{-8}$ to $2.3 \times 10^{-7}\,\mathrm{pc^{-3}}$. The inset shows the same distribution, together with the corresponding result from the \emph{ROSAT} Bright Survey (for the space density of long-period IPs only, from Pretorius et al.\ 2013), over-plotted as a dotted histogram.
}
 \label{fig:rhopdf}
\end{figure}

The distribution of $\rho_0$ values from the first simulation (Section~\ref{sec:calcobs}), normalized to give a probability distribution function, is shown in Fig.~\ref{fig:rhopdf}. The mode, median, and mean of the distribution are marked by solid vertical lines at $8.9 \times 10^{-8}$, $1.2 \times 10^{-7}$, and $2.3 \times 10^{-7}\,\mathrm{pc^{-3}}$. The dashed lines mark a 1-$\sigma$ confidence interval (the 16th and 84th percentile points of the distribution) between $7.4 \times 10^{-8}$ and $2.3 \times 10^{-7}\,\mathrm{pc^{-3}}$. Our best estimate of the mid-plane space density of long-period IPs is then $1^{+1}_{-0.5} \times 10^{-7}\,\mathrm{pc^{-3}}$.

The inset in Fig.~\ref{fig:rhopdf} shows the same distribution again (solid histogram), together with the probability distribution function of the space density of long-period IPs based on \emph{ROSAT} data (from \citealt{mcvrho}; dotted histogram, with a median at $1.7 \times 10^{-7}\,\mathrm{pc^{-3}}$).

\subsubsection{Upper limits on the space density of an undetected population}
\label{sec:limits}

\begin{figure}
 \includegraphics[width=84mm]{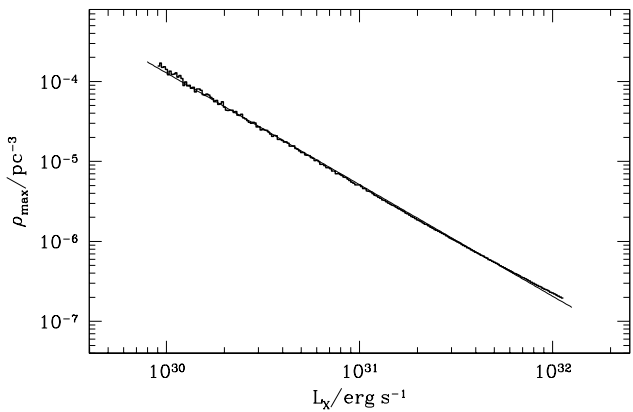} 
  \caption {The maximum allowed mid-plane space density of a hypothetical single-$L_X$ IP population, as a function of the 14--195~keV X-ray luminosity of that population. The bold histogram is the result of the simulation, and the fine line is a power-law fit (see text).}
 \label{fig:limit}
\end{figure}

The result of the second simulation (described in Section~\ref{sec:calclim}) is shown in Fig.~\ref{fig:limit}. This is the maximum allowed $\rho_0$ for a hypothetical undetected population of IPs, all with the same luminosity, as a function of that luminosity. The output of the simulation is shown as a bold histograms, and the fine curve is a fit to the data, given by
$$\rho_{max}= 5.15\times 10^{-6} (L_X/10^{31}\,\mathrm{erg\,s^{-1}})^{-1.40}\,\mathrm{pc^{-3}}.$$ 
This limit applies at luminosities below the values found for the observed sample (i.e.\ $L_X \la 10^{32} {\rm erg s^{-1}}$), since it is based on non-detection. As a pertinent example (see the discussion in the next section), its non-detection in our sample implies that a population of IPs at $L_X \sim 10^{31} {\rm erg s^{-1}}$ in the 14--195~keV band, can have a space density of at most $5\times 10^{-6}\,\mathrm{pc^{-3}}$ (or $\simeq 40 \times$ the space density of the observed population).

\section{Discussion}
\label{sec:discussion}

In a previous paper \citep{mcvrho} we reported space density estimates for IPs as well as polars, based on the \emph{ROSAT} Bright Survey. This soft X-ray survey was not very well suited to detecting objects with hard spectra, and the sample contained only 6 IPs. The IP $\rho$ measurement presented here is based on a larger sample, and in good agreement with our previous result (see the inset in Fig.~\ref{fig:rhopdf}). Here we find $1^{+1}_{-0.5} \times 10^{-7}\,\mathrm{pc^{-3}}$ for long-period IPs above $L_X \sim 10^{32} {\rm erg s^{-1}}$ in the BAT band, compared to $2^{+2}_{-0.9} \times 10^{-7}\,\mathrm{pc^{-3}}$ from the 4 long-period IPs in the \emph{ROSAT} sample.

Using the sample of all CVs at that time detected by \emph{INTEGRAL}, \cite{Revnivtsev08} reported a space density of $(1.5 \pm 0.6) \times 10^{-7}\,\mathrm{pc^{-3}}$ for long-period IPs. Although this is similar to the value presented here, it should be noted that the sample it was based on (besides not being flux-limited) included the dwarf nova SS Cyg, the asynchronous polar V1432 Aql, the polar IGR J14536-5522 \citep{Potter10}, and V709 Cas at a distance that has since been shown to be too small \citep{Thorstensen2010}. Those 4 systems account for more than half the total space density estimate of \cite{Revnivtsev08}, implying a significant systematic error.

\begin{figure}
 \includegraphics[width=84mm]{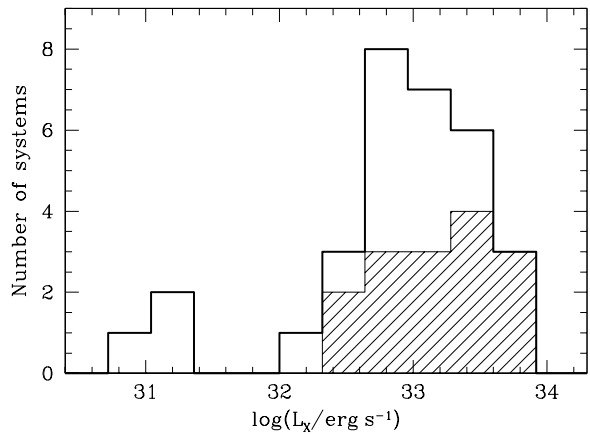} 
  \caption {The $L_X$ distribution of IPs detected in the \emph{Swift}-BAT 70 month survey. The main histogram shows the 31 securely classified IPs for which we were able to obtain distance estimates in the literature, or in the way described in Section~\ref{sec:dist}. The shaded histogram shows the 15 IPs in our sample (with V2731 Oph at its minimum allowed luminosity). Another three known IPs detected by BAT, V2069 Cyg, IGR J18173-2509 and SWIFT J2133.6+5105, are not included, because no useful constraints on their distances are available. The 3 systems at ${\rm log}(L_X/{\rm erg\,s^{-1}}) \sim 31$ are EX Hya, V1025 Cen, and DO Dra. The largest errors in $L_X$ are $\simeq 0.5$ dex (i.e.\ larger than the bin size), but several systems have much more precise estimates ($\le 0.08$ dex), namely EX Hya and DO Dra, in the fainter group, and GK Per, TV Col, and V1223 Sgr in the higher $L_X$ group.
}
 \label{fig:lxhist}
\end{figure}

In Fig.~\ref{fig:lxhist}, we show the $L_X$ distribution of all securely classified IPs detected in the \emph{Swift}-BAT 70 month survey, for which we are able to estimate luminosities. The majority have $\log (L_X/{\rm erg\,s^{-1}}) \ga 32$, but the distribution appears to be bimodal, with 3 systems (EX Hya, V1025 Cen, and DO Dra) found at ${\rm log}(L_X/{\rm erg\,s^{-1}}) \sim 31$.

Two of the 3 systems with X-ray luminosity below $10^{32} {\rm erg s^{-1}}$ have orbital periods below the period gap, where low accretion rates are expected. None of the other known short-period IPs (DW Cnc, HT Cam, V598 Peg, and CC Scl) has been detected by \emph{Swift}/BAT, and it is possible that they all belong to the low-luminosity population. 
The long-period IP at $L_X < 10^{32} {\rm erg s^{-1}}$, DO Dra, is a peculiar system in which the M-type secondary is readily visible \citep{Mukai90}, implying a relatively low accretion rate. A $0.8\,{\rm M_\odot}$ white dwarf accreting at $10^{-9}\,{\rm M_\odot/y}$ has a total accretion luminosity of $10^{34} {\rm erg\,s^{-1}}$. Considering the 14--195~keV to bolometric conversion factor, as well as the fraction of hard X-rays emitted downward and reprocessed into soft X-rays (roughly half), an accretion rate of this order is necessary to explain the high-luminosity IPs ($L_X \ga 10^{32} {\rm erg\,s^{-1}}$).

A single-temperature Bremsstrahlung fit to the BAT 70-month survey data of EX Hya, V1025 Cen, and DO Dra results in a temperature of 10.6, 15.1, and 15.2~keV, respectively. The only IPs in our core sample of 15 with similarly low temperatures are AO Psc (12.6~keV) and FO Aqr (15.2~keV). The low temperatures imply that Swift/BAT only detects a small fraction of the total hard X-ray luminosity of these IPs. Nevertheless, a 14--195 keV luminosity of $10^{31} {\rm erg s^{-1}}$ is incompatible with an accretion rate of $10^{-9}\,{\rm M_\odot/y}$. Moreover, detailed X-ray spectroscopy of HT Cam (likely a low-luminosity system; \citealt{deMartino05}) shows that it is much less absorbed than typical IPs, which may explain why several low-luminosity IPs were detected by \emph{ROSAT} but not by BAT.

We showed that a population of IPs at $L_X \sim 10^{31} {\rm erg s^{-1}}$ in the 14--195~keV band can have a space density as high as $5\times 10^{-6}\,\mathrm{pc^{-3}}$ (Section~\ref{sec:limits}). In that case, we expect (using the cumulative distribution function shown in figure 10 of \citealt{bat70month}) that the \emph{Swift}/BAT 70-month survey should contain $\simeq 12$ of these systems, compared to the 3 that are known at present. 
We can also find a (weak) lower limit on the space density of such low-luminosity IPs, by noting that all 3 detected systems are within 200~pc
[EX Hya and DO Dra have distances of $64.5 \pm 1.2$~pc and $155 \pm 35$~pc (\citealt{Beuermann03}; \citealt{MateoSzkodyGarnavich91}), and for V1025 Cen, we find a firm lower distance limit of 70~pc, and an estimate of roughly 110~pc]. 
The effective volume of the Galaxy at $d<200$~pc, conservatively assuming a larger scale-height of 260~pc, is $2.6\times 10^7\,{\rm pc^3}$. The low-luminosity IPs then have a space density of $>1\times 10^{-7}\,\mathrm{pc^{-3}}$. In other words, a low-luminosity IP population, likely at least as large as the the more frequently detected higher luminosity population, and possibly up to $\simeq 40 \times$ larger, exists. 

Almost all known IPs are found above the period gap (see e.g.\ figure~1 of \citealt{mcvrho}), where systems should spend only a short fraction of their lives, while most polars are short-period systems. This has led to the suggestion that IPs evolve into polars below the period gap (e.g.\ \citealt{ChanmugamRay84}). In \cite{mcvrho}, we showed that the space densities of long-period IPs and short-period polars are consistent with the quite simplistic model where all IPs become synchronized below the period gap, accounting for the entire short-period polar population. However, if the low-luminosity IP population is dominated by short-period systems (as appears to be the case), short-period IPs might be as intrinsically common as short-period polars (the space density of short-period polars is $7^{+5}_{-3}\times 10^{-7}\,\mathrm{pc^{-3}}$, if the high-state duty cycle is 0.5; \citealt{mcvrho}). Then another viable explanation for the fate of long-period IPs would be that they evolve into short-period IPs (see also \citealt{Southworth07}; \citealt{Norton08})\footnote{We expect the most reasonable explanation is that some long-periods IPs evolve into polars below the gap, while others become short-period IPs. However, currently available data cannot settle this question.}.


At the flux limit used to define our complete sample ($F_X=2.5 \times 10^{-11}\,\mathrm{erg\,cm^{-2}s^{-1}}$), \emph{Swift}-BAT has detected all $L_X = 10^{33} {\rm erg s^{-1}}$ IPs out to $\simeq 600$~pc, a distance at which non-magnetic CV samples are certainly very incomplete (e.g.\ \citealt{PretoriusKniggeKolb07}). However, even at a flux limit of $1.03 \times 10^{-11}\,\mathrm{erg\,cm^{-2}s^{-1}}$ (the level reached over 50\% of the sky by the \emph{Swift}-BAT 70-month survey) low-luminosity IPs ($L_X = 10^{31} {\rm erg s^{-1}}$) are only detected out to $<100$~pc. The all sky survey with \emph{eROSITA}, which is expected to reach an average sensitivity of $2 \times 10^{-13}\,\mathrm{erg\,cm^{-2}s^{-1}}$ in the 2--10~keV band \citep{Schwope12}, will therefore be important in the study of low-luminosity IPs in particular. {\it Gaia} will also be very important for our knowledge of CV space densities, since the error budget of our estimate is dominated by uncertainty in distances.

\section{Conclusions}
\label{sec:conclu}

To summarize, our main conclusions are:
\begin{enumerate}
\item Although many sources in the \emph{Swift}/BAT 70-month survey are still unidentified at longer wavelengths, it is possible to construct a complete, and still usefully large, IP sample from these data. We have obtained a complete flux-limited sample of 15 IPs, all above the period gap, by imposing the selection cuts $F_X > 2.5 \times 10^{-11}\,\mathrm{erg\,cm^{-2}s^{-1}}$ and $|b|>5^\circ$.
\item With the assumption that this sample is representative of the intrinsic population (in effect, assuming that IPs are brighter than roughly $10^{32}\,\mathrm{erg\,s^{-1}}$ in the 14--195~keV band), the mid-plane space density of long-period IPs is $1^{+1}_{-0.5} \times 10^{-7}\,\mathrm{pc^{-3}}$. This estimate is in good agreement with our previous measurement of $2^{+2}_{-0.9}\,\mathrm{pc^{-3}}$ for long-period IPs, from an independent data set \citep{mcvrho}.
\item We have also used the data to constrain the size of a fainter IP population that could exist without being included in our sample. A single-$L_X$ population of IPs with 14--195~keV luminosity below $\sim 10^{32}\,{\rm erg\,s^{-1}}$ must have a mid-plane space density less than $5.2\times 10^{-6} (L_X/10^{31}\,\mathrm{erg\,s^{-1}})^{-1.4}\,\mathrm{pc^{-3}}$.
\item Considering all IPs detected in the \emph{Swift}/BAT 70-month survey (rather than our brighter, complete sample of 15), we find evidence for a faint IP population, not included in the flux-limited sample. The $L_X$ distribution is bimodal, with peaks at $L_X \sim 10^{31}$ and $\sim 10^{33}\,{\rm erg\,s^{-1}}$.
\item The space density of the low-luminosity population is at least $10^{-7}\,\mathrm{pc^{-3}}$ (i.e.\ similar to the space density of the $L_X \ga 10^{32}\,\mathrm{erg\,s^{-1}}$ population to which our $\rho$ measurement applies) and possibly as high as $5\times 10^{-6}\,\mathrm{pc^{-3}}$ (or $\simeq 40 \times$ the space density of the bright population).
\item If the low-luminosity IP population is dominated by short-period systems, the above limits on its size are consistent with the simple model where all long-period IPs evolve into short-period IPs.
\end{enumerate}

\section*{Acknowledgements}
MLP is funded by a Marie Curie International Incoming Fellowship within the 7th European Community Framework Programme (grant no. PIIF-GA-2012-328716).

\bsp

\label{lastpage}

\end{document}